\documentclass[10pt, conference]{IEEEtran}
\IEEEoverridecommandlockouts

\usepackage[multiple]{footmisc}
\usepackage{algorithm}
\usepackage{algpseudocode}
\usepackage{amsmath}
\usepackage{array,booktabs}
\usepackage{balance}
\usepackage{cite}
\usepackage{colortbl}
\usepackage{comment}
\usepackage{float}
\usepackage{footnote}
\usepackage{listings}
\usepackage{makecell}
\usepackage{multirow}
\usepackage{pifont}
\usepackage{tabularray}
\usepackage{threeparttable}
\usepackage{xspace}
\usepackage{tcolorbox}

\definecolor{diffstart}{RGB}{173, 175, 177}
\definecolor{diffincl}{RGB}{0, 110, 0}
\definecolor{diffrem}{RGB}{255, 50, 0}
\definecolor{orange}{RGB}{255, 165, 0}
\definecolor{javapurple}{rgb}{0.5,0,0.35}

\newcommand{\revise}[1]{#1}
\newcommand{\revisecolor}{}

\newcommand{\find}[1]{
\begin{tcolorbox}[leftrule=0.3mm,rightrule=0.3mm, toprule=0.3mm,bottomrule=0.3mm,left=2pt,right=2pt,top=2pt,bottom=2pt]
{#1}
\end{tcolorbox}
}

\newcommand{\modelname}{FVF\xspace}

\begin{document}

\title{Similar but Patched Code Considered Harmful\\ \vspace{0.3cm}
\Large{The Impact of Similar but Patched Code on Recurring Vulnerability Detection\\  \vspace{0.05cm}
and How to Remove Them
}}

\makeatletter
\newcommand{\linebreakand}{
  \end{@IEEEauthorhalign}
  \hfill\mbox{}\par
  \mbox{}\hfill\begin{@IEEEauthorhalign}
}
\makeatother

\author{

\IEEEauthorblockN{Zixuan Tan\IEEEauthorrefmark{2},
Jiayuan Zhou\IEEEauthorrefmark{3},
Xing Hu\IEEEauthorrefmark{2}\IEEEauthorrefmark{1}\thanks{\IEEEauthorrefmark{1}Corresponding author},
Shengyi Pan\IEEEauthorrefmark{2},
Kui Liu\IEEEauthorrefmark{4},
Xin Xia\IEEEauthorrefmark{4}}
\IEEEauthorblockA{
\IEEEauthorrefmark{2}Zhejiang University, Hangzhou, Zhejiang, China \\
\IEEEauthorrefmark{3}Centre for Software Excellence, Huawei, Toronto, Canada \\
\IEEEauthorrefmark{4}Huawei, Hangzhou, Zhejiang, China \\
\{tanzixuan, xinghu, shengyi.pan\}@zju.edu.cn,
jiayuan.zhou1@huawei.com,
brucekuiliu@gmail.com,
xin.xia@acm.org
}
}

\maketitle

\begin{abstract}
Identifying recurring vulnerabilities is crucial for ensuring software security.
Clone-based techniques, while widely used, often generate many false alarms due to the existence of \textit{similar but patched}~(SBP) code, which is similar to vulnerable code but is not vulnerable due to having been patched.
Although the SBP code poses a great challenge to the effectiveness of existing approaches, it has not yet been well explored.

In this paper, we propose a {programming language agnostic} framework, Fixed Vulnerability Filter~(FVF), to identify and filter such SBP instances in vulnerability detection.
Different from existing studies that leverage function signatures, our approach analyzes code change histories to precisely pinpoint SBPs and consequently reduce false alarms.
Evaluation under practical scenarios confirms the effectiveness and precision of our approach.
\revise{Remarkably, FVF identifies and filters 65.1\% of false alarms from four vulnerability detection tools~(i.e., ReDeBug, VUDDY, MVP, and an elementary hash-based approach) without yielding false positives.}

We further apply FVF to 1,081 real-world software projects and construct a real-world SBP dataset containing 6,827 SBP functions.
Due to the SBP nature, the dataset can act as a strict benchmark to test the sensitivity of the vulnerability detection approach in distinguishing real vulnerabilities and SBPs.
Using this dataset, we demonstrate the ineffectiveness of four state-of-the-art deep learning-based vulnerability detection approaches.
{Our dataset can help developers make a more realistic evaluation of vulnerability detection approaches and also paves the way for further exploration of real-world SBP scenarios.}

\end{abstract}


\section{Introduction}
\label{sec:intro}
Code reuse is one of the most frequent activities in software development~\cite{lopes2017dejavu}.
By copying and pasting code snippets with or without modification, developers reuse existing code to improve the efficiency of programming.
However, vulnerabilities in the original code may also spread to downstream software.
For example, more than 60,000 open-source software projects are exposed to the vulnerability CVE-2017-12652~\cite{CVE-2017-12652} for the reuse of unsafe code snippets in the popular graphic library \textit{Libpng}~\cite{libpng_repo, reid2022extent}.
Due to poor software maintenance, these cloned \textit{similar vulnerabilities} introduced from the reuse process are difficult to detect~\cite{reid2022extent, tan2022understanding, jiang2020pdiff, farhang2019hey, nguyen2016automatic}.
Therefore, it is crucial for software maintainers to detect similar vulnerabilities in their codebases effectively.

\begin{figure}[tp]
\centerline{\includegraphics[width=\linewidth]{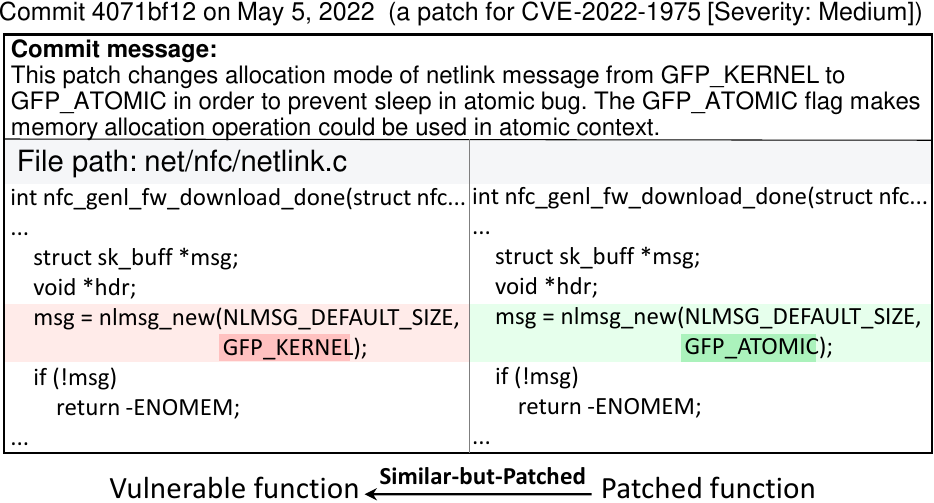}}
\vspace{-0.2cm}
\caption{An example showing the subtle difference between a vulnerable function (CVE-2022-1975) and the patched version~\cite{footnote_linux_4071bf12}.}
\label{fig:motivation_small_diff}
\vspace{-0.6cm}
\end{figure}

The code-clone-detection-based~(clone-based) approaches are commonly used to detect similar vulnerabilities~\cite{jang2012redebug, kim2017vuddy, li2016vulpecker, xiao2020mvp, reid2022extent}.
Generally, these techniques extract various signatures from vulnerable code and match similar code snippets as potential vulnerabilities.
However, due to the subtle differences between vulnerable code and its corresponding patched versions~\cite{xiao2020mvp}, it is a challenge for clone-based approaches to differentiate them effectively~(see Section \ref{sec:rq1} for our experimental result). This often leads to the misidentification of such \textbf{Similar-but-Patched}~(\textbf{SBP}) code as vulnerable, thus causing many false alarms.
Figure~\ref{fig:motivation_small_diff} shows an example of SBP code in the Linux kernel~\cite{footnote_linux_4071bf12} related to CVE-2022-1975~\cite{CVE-2022-1975}. The vulnerability was simply fixed by altering an argument of the \texttt{nlmsg\_new} function from \texttt{GFP\_KERNEL} to \texttt{GFP\_ATOMIC}). In this case, with only a single argument difference, the vulnerable function is closely similar to the patched version~(SBP).
Moreover, a piece of an SBP code could even exactly match a vulnerability. For example, if a vulnerability patch is reverted for various reasons (e.g., obsolescence or substitution with a better patch~\cite{shariffdeen2021automated}), the reverted code would become the same as the vulnerable code, but without maintaining its vulnerability since the vulnerability condition will not be triggered (see Figure~\ref{fig:rollback_type}). This poses a significant challenge for clone-based approaches that aim to determine vulnerability by only examining current code.

\revise{In practice, the inability to distinguish vulnerable and SBP code can lead to a large number of false alarms, requiring substantial human effort to manually verify the results, which is not always feasible and hinders the application of these approaches~\cite{meyer2014software}.}
MVP~\cite{xiao2020mvp}, proposed by Xiao et al., designed a function-level signature scheme to distinguish a vulnerability and an SBP. However, the proposed signature scheme is programming language-specific and lacks generalizability. Furthermore, MVP cannot handle the \textit{reverted type SBP} because its signature is identical to that of the vulnerability.

In this study, we propose a {programming language agnostic} framework, \textbf{FVF} (\textbf{F}ixed \textbf{V}ulnerability \textbf{F}ilter), to reduce false alarms in clone-based vulnerability detection by identifying and filtering the SBP code. The core idea behind FVF is to leverage code change histories to determine whether the detected potentially vulnerable code snippet has already been patched.
FVF works as a post-processing step for existing vulnerability detection approaches.
When a potentially vulnerable code snippet similar to a known vulnerability is detected, FVF queries the vulnerability feature database for a \textit{patch log}, which records the code change history of the vulnerability fix. It then retrieves the \textit{function change log} of the target code snippets.
\revise{Following existing recurring vulnerability detection approaches focusing on detecting function-level vulnerabilities, FVF generates the change and patch logs at the function level.}
If the \textit{patch log} is detected in the \textit{function change log}, it indicates that the potentially vulnerable code snippet has been patched previously (known as an SBP code snippet).

We evaluate the effectiveness of FVF in reducing false alarms~(i.e., SBP) in real-world scenarios. We adopt nine major versions of two popular open-source projects, namely the Linux kernel~\cite{footnote_kernel_org} and Redis~\cite{footnote_redis}, to evaluate how FVF can improve existing clone-based vulnerability detection approaches.
We employ \revise{four} popular clone-based vulnerability detection approaches, ReDeBug~\cite{jang2012redebug}, VUDDY~\cite{kim2017vuddy}, \revise{MVP~\cite{xiao2020mvp}}, and implement a simple hash-based approach, as baseline vulnerability detectors. The experimental results show that the overall False Alarm Rate~(FAR) for these detectors is \revise{76.2\%}, which is far from satisfactory and impractical. After applying FVF, the overall FAR is reduced to \revise{26.6\%}, with a significant improvement rate of \revise{49.6\%}.

We further analyze where and why FVF makes false predictions, including false positives of FVF and false negatives, and we find no false positives and summarize \revise{125} false negatives into two situations.
Based on the findings, we conduct a qualitative study on the characteristics of filtered SBPs that confuse clone-based vulnerability detection approaches. We categorize \revise{238} SBP code into three categories, which shed light on future research possibilities.

To evaluate the generalizability and scalability of FVF, we apply FVF on 1,081 historically vulnerable and popular open-source software~(OSS) projects written in C, C++, and Java programming languages.
In total, we collect 6,824 SBP functions and construct a dataset.
Using the dataset, we study the prevalence of the SBP code in the real world and observe that 40\% of OSS projects studied contain at least one instance of the SBP code.
The results confirm the prevalence of the SBP phenomenon, emphasizing the need to address the challenge.

Besides clone-based vulnerability detection techniques, deep learning-based (DL-based) techniques have gained promising performance~\cite{fu2022linevul, hanif2022vulberta, steenhoek2023empirical} in controlled lab environments.
However, prior studies~\cite{chakraborty2021reveal, steenhoek2023empirical} reveal that DL-based approaches sometimes leverage spurious features that are unrelated to the vulnerabilities, resulting in inferior performance in real-world scenarios.
Given the subtle differences between vulnerability and SBP, DL-based approaches may also fail to distinguish them, leading to a large number of false alarms.
Unfortunately, only a limited number of studies have considered SBP code and existing datasets such as Devign~\cite{zhou2019devign} and Big-Vul~\cite{fan2020bigvul} overlook the inclusion of SBP code.
As a result, DL-based approaches failed to learn SBP patterns during training, and evaluations become misaligned with real-world data distributions.
The impact of the SBP code on DL-based approaches is not well explored.

Using the collected dataset, we evaluate the performance of state-of-the-art DL-based vulnerability detection approaches. We select two token-based approaches (LineVul~\cite{fu2022linevul} and VulBERTa~\cite{hanif2022vulberta}) and two graph-based approaches (Devign~\cite{zhou2019devign} and IVDetect~\cite{li2021vulnerability}) for the study.
We use these approaches to detect vulnerabilities on the SBP dataset to assess the impact of SBP on DL-based vulnerability detection approaches.
The experimental results show that these approaches perform poorly on the dataset. All these approaches have a false alarm rate of more than 62\%. The token-based approaches mistakenly predict almost all SBP code as vulnerable, and the two graph-based approaches have a false alarm rate of 64.9\% and 62.9\%, respectively.
The results demonstrate the inability of these approaches to distinguish SBP from real vulnerabilities, thus emphasizing the discriminative effectiveness of our dataset.
Our dataset can help developers make a more realistic evaluation of existing vulnerability detection tools and also paves the way for further exploration of real-world SBP scenarios.

Our contributions are summarized as follows:
\begin{itemize}
    \item To the best of our knowledge, we are the first to systematically study the phenomenon of SBP and its impact on vulnerability detection.
    We find that while SBP code is prevalent in real-world scenarios, clone-based and DL-based vulnerability detection approaches are incapable of distinguishing SBP code, leading to a large number of false alarms in practice.
    \item We propose an effective framework, FVF, to identify SBP and help clone-based vulnerability detection approaches reduce false alarms. Experimental results show that FVF can significantly reduce false alarms of popular clone-based approaches such as VUDDY~\cite{kim2017vuddy} and ReDeBug~\cite{jang2012redebug}.
    \item We construct a real-world SBP dataset consisting of 6,824 SBP functions in three programming languages from 1,081 real-world projects using FVF, which can contribute to a more realistic evaluation of vulnerability detection tools.
    Our replication package can be accessed using the link~\cite{replication}.
\end{itemize}

{
The remainder of the paper is organized as follows:
Section \ref{sec:approach} describes the overview and design of our proposed framework.
Section \ref{sec:experiment} and Section \ref{sec:results} discuss the evaluation steps and results of {FVF}.
In Section \ref{sec:discussion}, we discuss other features of FVF, such as programming language agnostic, and the impact of SBP code on promising deep learning-based approaches.
In Section \ref{sec:threats}, we discuss the threats to the validity of our approach.
Section \ref{sec:related_work} summarizes related work in the field.
We conclude the paper in Section \ref{sec:conclusion}.
}

\section{FVF: The Proposed Approach}
\label{sec:approach}

\begin{figure*}[!t]
\begin{center}
\centerline{\includegraphics[width=.8\textwidth]{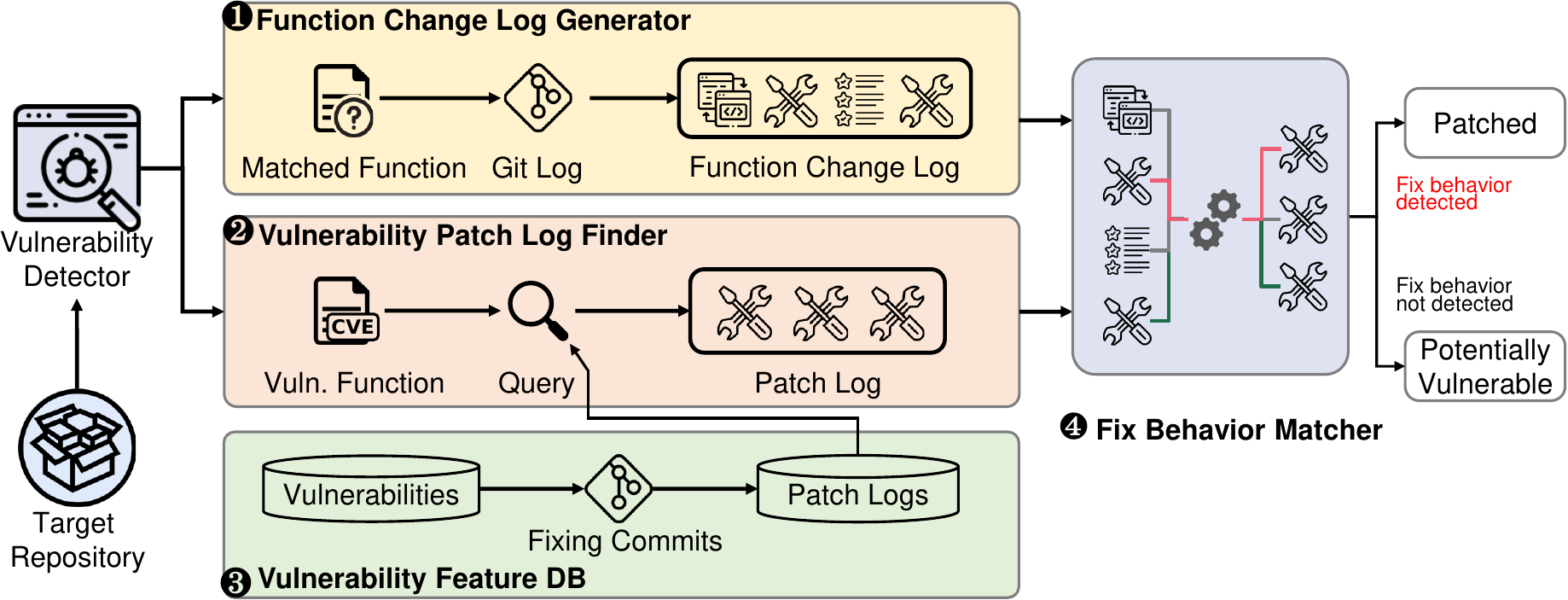}}
\vspace{-0.1cm}
\caption{Overview of FVF.}
\label{fig:overview}
\end{center}
\vspace{-0.9cm}
\end{figure*}

The goal of FVF is to enhance existing clone-based vulnerability detectors by reducing false alarms caused by already patched vulnerabilities~(i.e., SBP).
The core idea is to take the code changes that fix historical vulnerabilities as a reference, to check whether the detected vulnerable code has been fixed in the past.
In this section, we first present the overall framework \revise{and our design choices} of FVF, followed by the details of each component.

\begin{figure}[!t]
    \centering
    \includegraphics[width=\linewidth]{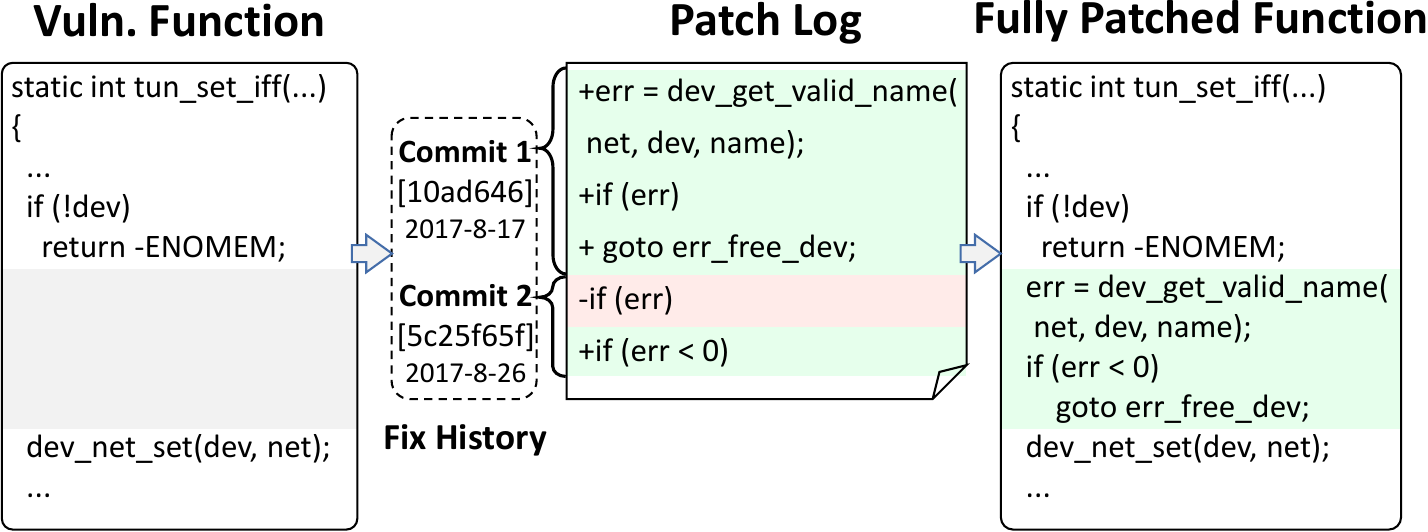}
    \vspace{-0.6cm}
    \caption{An example of generating the \textit{patch log} for function \texttt{tun\_set\_iff} and vulnerability CVE-2018-7191~\cite{CVE-2018-7191}. The \textit{patch log} contains two fixes~\cite{CVE-2018-7191_patch1, CVE-2018-7191_patch2}.}
    \label{fig:fix_log_generator}
    \vspace{-0.5cm}
\end{figure}

\subsection{Overall Framework}
\label{sec:overall_framework}
\revise{Following the existing recurring vulnerability detection approaches that focus on detecting function-level vulnerabilities, FVF also identifies SBP at the function level.}
The overall FVF framework, as shown in Figure \ref{fig:overview}, consists of four key components: \ding{182} \textbf{Function Change Log Generator}, \ding{183} \textbf{Vulnerability Patch Log Finder}, \ding{184} \textbf{Vulnerability Feature Database}, and \ding{185} \textbf{Fix Behavior Matcher}.

The vulnerability detector \revise{identifies vulnerabilities in the target repository and} produces the detected potentially vulnerable functions~(``Matched Function'' in Figure \ref{fig:overview}) to the \textbf{Function Change Log Generator} to generate a \textit{function change log}. Simultaneously, the detector produces the matched vulnerable function~(``Vuln. Function'' in Figure \ref{fig:overview}) to the \textbf{Vulnerability Patch Log Finder}, which queries the \textbf{Vulnerability Feature Database} to retrieve the \textit{patch logs}.
Finally, the \textbf{Fix Behavior Matcher} examines the \textit{function change log} for the presence of fix behaviors indicated in the \textit{patch logs}. If the same fix behavior is detected, the potentially vulnerable function is considered to have been previously fixed and is non-vulnerable (i.e., a false alarm). Otherwise, it remains potentially vulnerable and requires further review.

\subsection{Vulnerability Detector}
The Vulnerability Detector aims to detect recurring vulnerabilities in the target project.
When analyzing a target code repository, the detector produces two types of information: the matched function~(``Matched Function'') and the known vulnerable function~(``Vuln. Function'').
The matched function represents the potentially vulnerable function detected in the target code repository. The detector outputs the location (including the name, parameter definitions, and return value) of the matched function.
The vulnerable function refers to known vulnerable code that matches the detected potentially vulnerable function.
\revise{By requiring only the vulnerable functions and the matched similar functions, FVF supports diverse types of recurring vulnerability detectors, varying from string matching methods like ReDeBug~\cite{jang2012redebug}, signature-based methods like VUDDY~\cite{kim2017vuddy}, to more advanced slice-based techniques such as MVP~\cite{xiao2020mvp} and TRACER~\cite{kang2022tracer}.
This design ensures FVF's versatility and broad applicability.}
After detection, the matched and vulnerable functions are passed to the \textbf{Function Change Log Generator} and \textbf{Vulnerability Patch Log Finder}, respectively.

\subsection{Function Change Log Generator}
Given the matched function from the detector, the Function Change Log Generator retrieves the change history of the function in the version control system to build a \textit{function change log}.
\revise{A \textit{function change log} is a sequence of code changes, denoted as $\langle \text{fc}_1,\text{fc}_2,...,\text{fc}_m\rangle$, where $\text{fc}_i$ is a line-level code change on the function.}
The Function Change Log Generator first utilizes the \textit{git log} command~\cite{footnote_git_log} with the file path and function name as parameters to obtain the change history of the function.
Then all the change histories are concatenated in chronological order to generate the \textit{function change log}.

\revise{As retrieving the history in a large codebase can be very costly and slow down the whole process, we set a retrieval window to limit how far back in the version history we should look for changes.}
We consider the original vulnerability fix date to be the earliest date we should retrieve\revise{, as a cloned vulnerability is unlikely to be fixed earlier than the original one recorded in CVE.}
Additionally, we empirically set a threshold of 50 to limit the number of retrieval operations. This threshold is a configurable option, and we discuss efficiency and performance under different thresholds in Section~\ref{sec:discussion_efficiency}.

\revise{In practice, we find the retrieved history may be truncated in specific cases when the matched function is renamed or moved. Hence, we retrieve file-level change histories as a supplement. Specifically, when the function-level retrieval stops before reaching the window size threshold, we further retrieve the file change history and extract all modifications to the function.}

\subsection{Vulnerability Patch Log Finder and Feature Database (DB)}

\revise{The Vulnerability Patch Log Finder retrieves the appropriate \textit{patch log} for the vulnerable function.
It first gets the necessary information (e.g., the signatures and line numbers) about the vulnerable functions from the vulnerability detector. Then, it queries the Vulnerability Feature Database with CVE ID and vulnerable function signature for the \textit{patch log}.}

\revise{A \textit{patch log} records all the fix actions, represented as a sequence of line-level code diffs concatenated from a series of patches on the function, chronologically.}
Figure~\ref{fig:fix_log_generator} shows an example of generating a \textit{patch log} for the function \texttt{tun\_set\_iff} and the vulnerability CVE-2018-7191~\cite{CVE-2018-7191}.
If multiple fixes have been applied to the same function, the last post-fix version is considered the fully fixed version. Thus, the \textit{patch log} records each patch diff chronologically from the first patch to the last patch, making it naturally support multi-patch scenarios.

\revise{
The Vulnerability Feature Database stores the \textit{patch logs} of all the disclosed vulnerabilities for the Vulnerability Patch Log Finder to query.
Specifically,  we collect the patch information of the disclosed vulnerabilities from vulnerability databases (e.g., NVD) and process to generate \textit{patch logs} for each vulnerable function modified in the vulnerability patch.
Besides, the database is continuously updated with the patch log of newly disclosed vulnerabilities, ensuring that the latest vulnerability information is always available for querying.
}

\subsection{Fix Behavior Matcher}
Given a \textit{function change log} and a \textit{patch log}, the Fix Behavior Matcher determines if the \textit{patch log} is already contained in the \textit{function change log}.
If the condition is true, it means the function has fix behaviors in the past, suggesting the vulnerability has already been fixed. Hence, the detected result is an SBP case and is considered a false alarm.

Formally, for {function change log} $\text{FC\_Log} = \langle \text{fc}_1,...,\text{fc}_m\rangle$ and {patch log} $\text{Pat\_Log} = \langle \text{pc}_1,...,\text{pc}_n\rangle$,
if there exists a set of indexes where ${1\leq i_1<i_2<...<i_n\leq m}$ such that $\textit{Sim}(\text{pc}_{j}, \text{fc}_{i_j}) \geq \text{Threshold}$ for $1\leq j\leq n$, we consider $\text{Pat\_Log}$ to be contained in $\text{FC\_Log}$.
If it is contained, it indicates that the potentially vulnerable function detected is an SBP one. Therefore, this detection result is a false alarm.
Instead of direct string matching, we employ similarity to make FVF more robust when the cloned version has different literal representations, such as using different variable names or function names, etc. We denote \textit{Sim} as a similarity calculator and calculate the BLEU-2~\cite{papineni2002bleu} score.
The BLEU-2 score quantifies the similarity of 2-grams, i.e., consecutive pairs of words, making it especially applicable in scenarios where only the identifier name varies.

Since FVF is only interested in the subsequence of the \textit{patch log} during the matching process, the code changes in the function history that are not related to the patch do not affect the result.

\section{Experiments}
\label{sec:experiment}

\noindent In this paper, we aim to answer the following research questions:

\noindent\textbf{RQ1: How effective is FVF in identifying SBP and reducing false alarms?}
Clone-based vulnerability detection approaches are often not practical as they struggle to distinguish vulnerabilities and SBP code, {resulting in a large amount of false alarms~\cite{jang2012redebug, xiao2020mvp}.}
The goal of this RQ is to evaluate the false alarm rate of the existing clone-based vulnerability detection approaches and then the effectiveness of FVF in reducing false alarms.

\noindent\textbf{{RQ2: What are the false predictions of FVF in {identifying false alarms?}}}
In this RQ, we look into the details of when FVF fails in identifying SBP code.
Especially, we investigate the FPs (i.e., real recurring vulnerabilities that are incorrectly identified as SBP code snippets) and FNs (i.e., real false alarms that are not identified) of FVF.

\noindent\textbf{RQ3: What are the characteristics of filtered SBP code snippets?}
In this RQ, we further conduct a qualitative study on the filtered SBP code snippets, to gain empirical insights on their characteristics, including the categories of SBP code snippets and the reason for each category.

\subsection{Data Collection and Preprocessing}
\label{sec:experiment_data}

\begin{figure}
    \centering
    \includegraphics[width=0.8\linewidth]{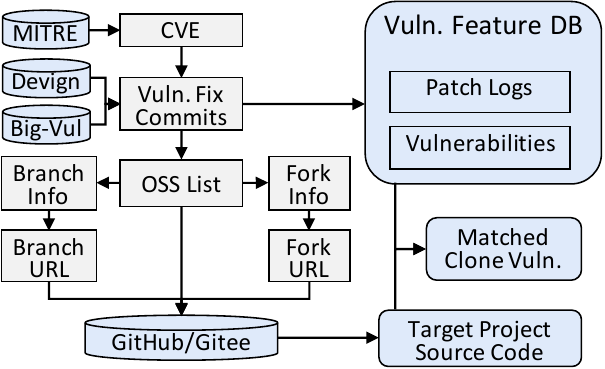}
    \vspace{-0.1cm}
    \caption{An overview of the data collection approach.}
    \label{fig:data-collection}
    \vspace{-0.5cm}
\end{figure}

\noindent Our experiment data include two parts: 1) Vulnerability feature database and 2) Target project source code.
The vulnerability feature database contains information on existing vulnerabilities, which is used by FVF. The target project is the project from which FVF identifies SBP code snippets.
Figure \ref{fig:data-collection} illustrates the workflow of our data collection process. We describe the process in detail below.

\subsubsection{Vulnerability feature database} The database includes two main parts: vulnerability information and \textit{patch logs}.
Initially, vulnerability information is collected from multiple reliable sources, followed by the generation of \textit{patch logs}. The process of vulnerability feature database construction is outlined as follows:

\noindent\textbf{Step 1: Collecting Vulnerability Fix Commits~(VFCs).}
Vulnerability fix commits are commits in the version control system that fix vulnerabilities.
We collect VFCs from two existing vulnerability datasets, namely Devign~\cite{zhou2019devign} and Big-Vul~\cite{fan2020bigvul}, resulting in 6,611 and 3,746 VFCs, respectively.
Additionally, we extract more VFCs from the reference links of the CVE records in the authoritative vulnerability information source, the MITRE CVE database~\cite{footnote_mitre}.
Note that we only keep VFCs modifying C/C++ source files.
After deduplication, we collect a total of 15,636 VFCs distributed across 978 projects.

\noindent \textbf{Step 2: Cleaning Noisy Data.}
It is important to note that not all changes within a VFC are related to fixing vulnerabilities. Some commits contain multiple intentions~(e.g., code refactoring), which could introduce additional noise.
We conduct a statistical analysis on the number of files modified in each commit within the dataset and find that the 99th percentile of the number of modified files per vulnerability fix is 10. Accordingly, VFCs that modify more than 10 files are excluded.
We additionally exclude noisy commits that only revise comments or adjust white spaces.

\noindent \textbf{Step 3: Extracting Vulnerability-Relevant Functions.}
We extract both pre- and post-commit versions of modified functions from each VFC using PyDriller~\cite{PyDriller}.
Following previous studies~\cite{zhou2019devign, fan2020bigvul}, the pre-commit version is labeled as vulnerable, while the post-commit version is considered patched.
In the case of multiple VFCs for a single vulnerability, only the post-commit version of the latest VFC is considered patched; otherwise, it remains vulnerable.

\noindent\textbf{Step 4: Generating Patch Log.}
Figure~\ref{fig:fix_log_generator} shows an example of generating a \textit{patch log}.
A \textit{patch log} is a sequence of modification lines beginning with `+' or `-', which records every code change made to the vulnerable function, from the initial VFC to the latest one chronologically. Specifically, the modification lines related to the vulnerable function in each VFC are concatenated into a unified sequence to build a \textit{patch log}.
In total, we collect 35,319 vulnerable functions, and 18,074 \textit{patch logs} are generated.

\begin{table}
    \caption{Details of selected branches of target OSS.}
    \vspace{-0.2cm}
    \centering
    \resizebox{.9\linewidth}{!}{
    \begin{tabular}{llllrr}
    \toprule
    \textbf{OSS}                                   & \textbf{Branch} & \textbf{Abbr.}     & \textbf{Status}      & \textbf{Function} & \textbf{LoC} \\ \midrule
    \multirow{3}{*}{\textbf{\makecell[vl]{Linux\\Kernel}}} & 6.3-rc5  &  L.6.3  & Mainline &     657K    &   16M   \\
                                           & 6.2.9    &  L.6.2  & Stable   &     656K    &   16M   \\
                                           & 5.15.105 &  L.5.15 & LTS      &     616K    &   15M   \\ \midrule
    \multirow{2}{*}{\textbf{Redis}}       & 7.0.10  & R.7    & Stable      &     7K      &   154K   \\
                                          & 5.0.14  & R.5    & LTS         &     5K      &   105K    \\

                             \bottomrule
    \end{tabular}
    }
    \label{table:oss_upstream}
    \vspace{-0.3cm}
    \end{table}

\begin{table}
    \caption{Details of selected forks of target OSS.}
    \vspace{-0.2cm}
    \resizebox{\linewidth}{!}{
    \begin{tabular}{lllrrr}
    \toprule
    \textbf{\makecell[vl]{OSS}}                               & \textbf{\makecell[vl]{Downstream \\ Fork}}    & \textbf{\makecell[vl]{Abbr.}}             & \textbf{\makecell[vr]{Commits \\ Ahead}} & \textbf{\makecell[vr]{Commits \\ Behind}} & \textbf{\makecell[vl]{Last \\ Update}} \\ \midrule
    \multirow{3}{*}{\textbf{\makecell[vl]{Linux\\Kernel}}}   & Asahi Linux                                    & L.A  & 2,230                 & 32,401                  & 2022-12                  \\
                                      & \mbox{Linux~Kernel~Lib.}                           & L.L  & 975                   & 80,518                  & 2023-02                  \\
                                      & O.H. Linux~Kernel 5.10 & L.O  & 11,076                    & 273,473                     & 2023-04                  \\ \midrule
        \textbf{Redis}                               & Birdisle                                       & R.B  & 666                   & 4,521                   & 2023-04                  \\
                                      \bottomrule
    \end{tabular}
    }
    \label{table:oss_fork}
    \vspace{-0.5cm}
    \end{table}

\subsubsection{Target project source code}
We follow the existing clone-based vulnerability detection approaches~\cite{kim2017vuddy, xiao2020mvp}
to adopt the Linux kernel~\cite{footnote_kernel_org} and Redis~\cite{footnote_redis}, two widely used OSS projects \revise{with extensive branches and forks}, as our target projects.
Linux is a widely used operating system kernel, while Redis is a popular key-value database system extensively used by companies.
{Then we further collect the source code of the branches and forks of the target OSS projects.}

Table~\ref{table:oss_upstream} provides details of the branches we select for the Linux kernel and Redis.
For Linux, we select the presently latest version~(tag: \textbf{\textit{6.3-rc5}}) on the master branch, a stable release~(tag: \textbf{\textit{6.2.9}}), and a long-term support version~(tag: \textbf{\textit{5.15.105}}).
For Redis, we select one stable version~(tag: \textbf{\textit{7.0.10}}) and one long-term support branch~(tag: \textbf{\textit{5.0.14}}).

In supplement of branches, we select OSS forks that make customization and still actively commit in the most recent six months.
Details of the forks for the Linux kernel and Redis are outlined in Table~\ref{table:oss_fork}.
The term \textit{Commits Ahead} denotes the number of exclusive commits in the forked version, indicating the evolving history of the project. \textit{Commits Behind} denotes the number of upstream commits not yet to be incorporated into the fork, indicating the level of outdatedness.
For Linux, three downstream forks are chosen:
\textit{\textbf{Asahi Linux}}~\cite{footnote_asahilinux}, which aims to adapt the Linux kernel to Apple silicon Mac computers.
\textit{\textbf{Linux Kernel Library}}~\cite{footnote_lkl_linux}, focusing on reusing Linux kernel code as a library for user-level applications.
\textit{\textbf{OpenHarmony Linux kernel 5.10}}~\cite{footnote_openharmony}, a customized Linux kernel maintained by OpenHarmony~(O.H.), an open-source project for the Internet of Things~(IoT) devices.
Regarding Redis, one fork is selected:
\textit{\textbf{Birdisle}}~\cite{footnote_birdisle}, a modified Redis version that operates as a library within another process.

\subsection{Baselines}
\label{sec:baselines}

\noindent \revise{We evaluate our approach with four baselines, including three well-known vulnerable code clone detection approaches (i.e., ReDeBug, VUDDY, and MVP) and a hash-based approach.}

\noindent\textbf{ReDeBug}~\cite{jang2012redebug} extracts vulnerable signatures from vulnerability patches and leverages a pattern-matching approach to detect unpatched code clones. ReDeBug takes the deleted lines and the surrounding context lines in the patch file to generate the vulnerable signature.

\noindent\textbf{VUDDY}~\cite{kim2017vuddy} employs abstraction and normalization techniques to generate coarse-grained and vulnerability-preserving function signatures for vulnerable code clone detection.

\noindent\revise{\textbf{MVP}~\cite{xiao2020mvp} employs program slicing techniques to extract vulnerability and patch signatures, identifying recurring vulnerabilities that match the vulnerability signatures while not matching the patch signatures.}

\noindent\textbf{Hash-based} is a simple function-matching approach that aims to evaluate the effectiveness of FVF even for such a trivial approach. This approach generates a hash ID for a given function as the signature of the function. For a given function, the approach compares the signature with all signatures of known vulnerable functions. If two signatures are matched, the function is considered to be a recurring vulnerability. We simply abstract the names and parameters of functions to obtain a slight generalization ability.

\subsection{Test Scenarios}
\label{sec:test_scenarios}
\noindent We evaluate the false alarm rate of baselines before and after applying our proposed approach in detecting vulnerabilities of target projects.
Specifically, we run the baseline vulnerability detection tools on each target project, then apply FVF on the detected potential vulnerabilities, to identify false alarms caused by SBP.

Due to the inherent challenge of identifying vulnerabilities, there is no definitive ground truth on all possible vulnerabilities. To establish a fair and reliable ground truth, we engage a team of three security professionals, each with at least three years of experience in software security, to manually validate each potential vulnerability detected by the baselines.

\revise{The validation process starts with two security professionals independently labeling the potential vulnerabilities identified by the approach.
Specifically, each professional is asked to first review the detailed disclosed information of the source vulnerability, including the CVE description and the vulnerability patch, to gain a comprehensive understanding of the vulnerability root cause.
Then, based on such understanding of the source vulnerability, the professional evaluates whether the detected vulnerability is truly a recurring vulnerability.
Subsequently, the labeling results of the two professionals are cross-verified. We employ Cohen's Kappa~\cite{mchugh2012interrater} to measure the inter-rater reliability, resulting in a value of 0.79, which indicates a substantial agreement.
To increase reliability, the remaining discrepancies are resolved by the third professional using a similar review approach. The professionals have ample time for thorough inspections and are well compensated for their effort and expertise.
}

\subsection{Evaluation Metrics}
\label{sec:evaluation_metrics}
\noindent As our goal is to minimize the number of false alarms in clone-based vulnerability detection, we evaluate the baselines and the improvement made by FVF through three \revise{false-alarm-centric} metrics: the number of false alarms, the false alarm rate \revise{(equivalent to false alarm precision)}, and the improvement rate.
\revise{We do not calculate recall since we do not have the ground truth of detected negatives for the baseline detector.}

\noindent\textbf{False Alarm (FA)} represents the number of false alarms produced by the baselines. A higher FA indicates a lower detection accuracy of an approach.

\noindent\textbf{False Alarm Rate (FAR)} is the ratio of false alarms to the total number of predicted positives. FAR reflects the accuracy of positive predictions made by the detector. A higher FAR indicates lower quality in detection performance.

\noindent\textbf{Improvement Rate (IR)} is the observed improvement rate of false alarms~(i.e., the reduction in FAR) before and after applying our proposed approach. A higher IR indicates a more substantial enhancement brought by FVF.

\subsection{Implementation Details}
\label{sec:experiment_setup}

\noindent We implement FVF in 1.5k lines of Python code. Our experiments are performed on an Ubuntu 20.04 server, with 2 Intel Xeon Gold 6226R CPUs and 256G RAM.
For ReDeBug~\cite{jang2012redebug}, we follow the original configuration to set the length of n-gram to 4.
For VUDDY~\cite{kim2017vuddy}, we use the online platform offered on their official website~\cite{footnote_iotcube}. We first generate signatures locally and then upload them to the platform for detection.
\revise{For MVP~\cite{xiao2020mvp}, we reimplement the algorithm strictly according to the methodology in the original paper and use the recommended parameters as the original paper as well.
}
For the hash-based approach, we leverage the MD5 algorithm to generate hash IDs and compare the hash IDs.

\section{Experimental Results}
\label{sec:results}

\subsection{RQ1: The Effectiveness of FVF in Identifying SBP and Reducing False Alarms}
\label{sec:rq1}

\begin{table}
    \revisecolor
    \caption{The performance of baselines in detecting vulnerabilities in Linux and Redis before/after adopting FVF (TP: True positive, FA: False alarm, FAR: False alarm rate, IR: Improvement rate.)}
    \begin{tabular}{p{0.45cm}p{1.05cm}rrrrrr}
    \toprule
    \multirow{2}{*}{\textbf{\makecell[vl]{Proj.\\Abbr.}}}
                                        & \textbf{Approach} & \textbf{TP} & \multicolumn{2}{l}{\textbf{Original Perf.}} & \multicolumn{3}{l}{\textbf{After \modelname}} \\%
    \cmidrule(lr){4-5} \cmidrule(lr){6-8}
                                        &            &             & \textbf{FA} & \textbf{\mbox{FAR(\%)}} & \textbf{FA} & \textbf{\mbox{FAR(\%)}}   &    \textbf{IR(\%)}     \\ \midrule
    \multirow{3}{*}{\textbf{L.6.3}} & \mbox{Hash-based} &   0  &  23   &  100.0   & \cellcolor[HTML]{E6E6E6} 2   &  \cellcolor[HTML]{E6E6E6} 13.0   & \cellcolor[HTML]{E6E6E6} 87.0  \\
                                        & VUDDY      &      0  &  13   &  100.0   & \cellcolor[HTML]{E6E6E6}  2  & \cellcolor[HTML]{E6E6E6}  15.4   & \cellcolor[HTML]{E6E6E6} 84.6  \\
                                        & ReDeBug    &      0  &  90   &  100.0   & \cellcolor[HTML]{E6E6E6} 38  & \cellcolor[HTML]{E6E6E6}  42.2   & \cellcolor[HTML]{E6E6E6} 57.8  \\
                                        & MVP       &      0  &  149   &  100.0   & \cellcolor[HTML]{E6E6E6}  23  & \cellcolor[HTML]{E6E6E6}  15.4  & \cellcolor[HTML]{E6E6E6}  84.6 \\ \midrule
    \multirow{3}{*}{\textbf{L.6.2}} & \mbox{Hash-based} &   0  &  24   &  100.0   & \cellcolor[HTML]{E6E6E6} 3   & \cellcolor[HTML]{E6E6E6}  12.5   & \cellcolor[HTML]{E6E6E6} 87.5  \\
                                        & VUDDY      &      0  &  13   &  100.0   & \cellcolor[HTML]{E6E6E6}  2  & \cellcolor[HTML]{E6E6E6}  15.4   & \cellcolor[HTML]{E6E6E6} 84.6  \\
                                        & ReDeBug    &      0  &  91   &  100.0   & \cellcolor[HTML]{E6E6E6} 38  & \cellcolor[HTML]{E6E6E6}  41.8   & \cellcolor[HTML]{E6E6E6} 58.2  \\
                                        & MVP       &      0  &  153   &  100.0   & \cellcolor[HTML]{E6E6E6}  23  & \cellcolor[HTML]{E6E6E6}  15.0  & \cellcolor[HTML]{E6E6E6} 85.0  \\ \midrule
    \multirow{3}{*}{\textbf{L5.15}} & \mbox{Hash-based} &   0  &  29   &  100.0   & \cellcolor[HTML]{E6E6E6} 4   & \cellcolor[HTML]{E6E6E6}  13.8   & \cellcolor[HTML]{E6E6E6} 86.2  \\
                                        & VUDDY      &      0  &  16   &  100.0   & \cellcolor[HTML]{E6E6E6}  3  & \cellcolor[HTML]{E6E6E6}  18.8   & \cellcolor[HTML]{E6E6E6} 81.3  \\
                                        & ReDeBug    &      0  &  93   &  100.0   & \cellcolor[HTML]{E6E6E6} 40  & \cellcolor[HTML]{E6E6E6}  43.0   & \cellcolor[HTML]{E6E6E6} 57.0  \\
                                        & MVP       &      37  &  165   &  81.7   & \cellcolor[HTML]{E6E6E6} 28  & \cellcolor[HTML]{E6E6E6}  13.9   & \cellcolor[HTML]{E6E6E6} 67.8  \\ \midrule
    \multirow{3}{*}{\textbf{L.A}}   & \mbox{Hash-based} &   0  &  27   &  100.0   & \cellcolor[HTML]{E6E6E6} 3   & \cellcolor[HTML]{E6E6E6}  11.1   & \cellcolor[HTML]{E6E6E6} 88.9  \\
                                        & VUDDY      &      0  &  15   &  100.0   & \cellcolor[HTML]{E6E6E6}  2  & \cellcolor[HTML]{E6E6E6}  13.3   & \cellcolor[HTML]{E6E6E6} 86.7  \\
                                        & ReDeBug    &      0  &  92   &  100.0   & \cellcolor[HTML]{E6E6E6} 38  & \cellcolor[HTML]{E6E6E6}  41.3   & \cellcolor[HTML]{E6E6E6} 58.7  \\
                                        & MVP       &      0  &  149   &  100.0   & \cellcolor[HTML]{E6E6E6}  23  & \cellcolor[HTML]{E6E6E6}  15.4   & \cellcolor[HTML]{E6E6E6} 84.6  \\ \midrule
    \multirow{3}{*}{\textbf{L.L}}   & \mbox{Hash-based} &  21  &  27   &   56.3   & \cellcolor[HTML]{E6E6E6} 3   & \cellcolor[HTML]{E6E6E6}   6.3   & \cellcolor[HTML]{E6E6E6} 50.0  \\
                                        & VUDDY      &      3  &  16   &   84.2   & \cellcolor[HTML]{E6E6E6} 2   & \cellcolor[HTML]{E6E6E6}  10.5   & \cellcolor[HTML]{E6E6E6} 73.7  \\
                                        & ReDeBug    &     18  &  95   &   84.1   & \cellcolor[HTML]{E6E6E6} 40  & \cellcolor[HTML]{E6E6E6}  35.4   & \cellcolor[HTML]{E6E6E6} 48.7  \\
                                        & MVP       &     41  &   149   &   78.4   & \cellcolor[HTML]{E6E6E6} 21   & \cellcolor[HTML]{E6E6E6}  11.1   & \cellcolor[HTML]{E6E6E6}  67.3  \\ \midrule
    \multirow{3}{*}{\textbf{L.O}}   & \mbox{Hash-based} &   0  &  29   &   100.0  & \cellcolor[HTML]{E6E6E6}  4  & \cellcolor[HTML]{E6E6E6}    13.8   & \cellcolor[HTML]{E6E6E6} 86.2  \\
                                        & VUDDY      &      0  &  16   &   100.0  & \cellcolor[HTML]{E6E6E6}  3  & \cellcolor[HTML]{E6E6E6}    18.8   & \cellcolor[HTML]{E6E6E6} 81.3  \\
                                        & ReDeBug    &      0  & 112   &   100.0  & \cellcolor[HTML]{E6E6E6} 43  & \cellcolor[HTML]{E6E6E6}    38.4   & \cellcolor[HTML]{E6E6E6} 61.6  \\
                                        & MVP       &      47  &  157   &   77.0  & \cellcolor[HTML]{E6E6E6} 27  & \cellcolor[HTML]{E6E6E6}    13.2   & \cellcolor[HTML]{E6E6E6} 63.8  \\ \midrule
    \multirow{3}{*}{\textbf{R.7}}   & \mbox{Hash-based} &   0  &   5   &   100.0  & \cellcolor[HTML]{E6E6E6}  0  & \cellcolor[HTML]{E6E6E6}     0.0     & \cellcolor[HTML]{E6E6E6} 100.0  \\
                                        & VUDDY      &      0  &   3   &   100.0  & \cellcolor[HTML]{E6E6E6}  0  & \cellcolor[HTML]{E6E6E6}     0.0    & \cellcolor[HTML]{E6E6E6} 100.0  \\
                                        & ReDeBug    &      0  &   5   &   100.0  & \cellcolor[HTML]{E6E6E6}  0  & \cellcolor[HTML]{E6E6E6}     0.0    & \cellcolor[HTML]{E6E6E6} 100.0  \\
                                        & MVP        &      0  &  13   &  100.0   & \cellcolor[HTML]{E6E6E6}  0  & \cellcolor[HTML]{E6E6E6}     0.0    & \cellcolor[HTML]{E6E6E6} 100.0  \\ \midrule
    \multirow{3}{*}{\textbf{R.5}}   & \mbox{Hash-based} &   7  &   3   &   30.0   & \cellcolor[HTML]{E6E6E6}  0  & \cellcolor[HTML]{E6E6E6}     0.0      & \cellcolor[HTML]{E6E6E6} 30.0        \\
                                        & VUDDY      &      2  &   2   &   50.0   & \cellcolor[HTML]{E6E6E6}  0  & \cellcolor[HTML]{E6E6E6}     0.0      & \cellcolor[HTML]{E6E6E6} 50.0    \\
                                        & ReDeBug    &      8  &   3   &   27.3   & \cellcolor[HTML]{E6E6E6}  0  & \cellcolor[HTML]{E6E6E6}     0.0      & \cellcolor[HTML]{E6E6E6} 27.3    \\
                                        & MVP        &      4  &   10  &    71.4   & \cellcolor[HTML]{E6E6E6}  0  & \cellcolor[HTML]{E6E6E6}    0.0     & \cellcolor[HTML]{E6E6E6}  71.4    \\ \midrule
    \multirow{3}{*}{\textbf{R.B}}   & \mbox{Hash-based} &   7  &  3    &   30.0   & \cellcolor[HTML]{E6E6E6}  0  & \cellcolor[HTML]{E6E6E6}    0.0      & \cellcolor[HTML]{E6E6E6} 30.0        \\
                                        & VUDDY      &      2  &  2    &   50.0   & \cellcolor[HTML]{E6E6E6}  0  & \cellcolor[HTML]{E6E6E6}    0.0      & \cellcolor[HTML]{E6E6E6} 50.0        \\
                                        & ReDeBug    &      8  &  3    &   27.3   & \cellcolor[HTML]{E6E6E6}  0  & \cellcolor[HTML]{E6E6E6}    0.0      & \cellcolor[HTML]{E6E6E6} 27.3      \\
                                        & MVP        &     49  & 4  &  7.5  & \cellcolor[HTML]{E6E6E6} 0   & \cellcolor[HTML]{E6E6E6}  0.0   & \cellcolor[HTML]{E6E6E6}  7.5    \\ \midrule
    \textbf{Total}                      &  -         &     112 & 358   &    76.2      & \cellcolor[HTML]{E6E6E6}  125 & \cellcolor[HTML]{E6E6E6}    26.6    & \cellcolor[HTML]{E6E6E6}  49.6       \\
    \bottomrule
    \end{tabular}
    \label{table:rq_code_clone_perf}
    \vspace{-0.7cm}
    \end{table}

\noindent In total, the \revise{four} baselines produce \revise{470} potential vulnerabilities after deduplication.
Three security professionals with at least three years of security experience manually verify all potential vulnerabilities produced.
{The manual verification process is described in detail in Section \ref{sec:test_scenarios}.}
After verification, \revise{112} potential vulnerabilities are confirmed as real (i.e., true positives, TP) in total.
The results are presented in Table~\ref{table:rq_code_clone_perf}.
For the Linux kernel, all \revise{75} TPs are detected in the L.L project, indicating a poor security maintenance status.
Regarding Redis, the latest version (R.7) is well-maintained but missing fixes are found in R.5 and R.B.

Among all approaches, the hash-based model is the most trivial.
For the Linux kernel, the detection results of the hash-based model are 100\% false alarms except for the LKL (L.L) project.
After applying FVF, the FAR decreases and ranges from 6.3\% to 13.8\%. The IR ranges from 50.0\% to 88.9\%.
For Redis, the FAR ranges from 30.0\% to 100.0\%. After applying FVF, all the false alarms are identified and filtered.

VUDDY generates more coarse-grained function-level signatures than the hash-based model to achieve better performance.
However, VUDDY performed the worst among the baselines.
For the Linux kernel, all the results reported by VUDDY are false alarms, with the exception of the LKL project, which exhibits a notably high FAR of 84.2\%.
For Redis, the FAR ranges from 50.0\% to 100.0\%.
One possible explanation for the high FA could be the quality of the vulnerability database. VUDDY relies on its \revise{private online} database for detection, which may be outdated and not comprehensive.
After enhancing VUDDY by applying FVF, the FAR in Linux is effectively reduced, dropping by 73.7\%-86.7\%. For Redis, all false alarms are identified and filtered.

ReDeBug constructs vulnerability signatures using partial information on vulnerabilities to identify cloned vulnerabilities, which makes ReDeBug more robust but also introduces more false alarms.
For the Linux kernel, ReDeBug produces 90 to 112 false alarms, with a FAR of 100.0\% across all versions except for the LKL, which is 84.1\%.
For Redis, the FAR ranges from 27.3\% to 100.0\%.
After applying FVF, the FARs on the Linux kernel projects decreased significantly, dropping by 48.7\% to 61.6\%. Furthermore, FVF filtered out all false alarms on Redis.

\revise{MVP leverages code-slicing techniques on the vulnerable and patched function to generate vulnerability signatures, which makes the signature more comprehensive in detecting potential similar vulnerabilities. However, it still predicts many false alarms due to not considering reverted-type SBPs. For the six versions of the Linux kernel, MVP predicts 149 to 214 similar vulnerabilities and the false alarm rate is between 77\% and 100\%. For Redis, MVP predicts 13 to 53 similar vulnerabilities and the false alarm rate is between 7.5\% and 100\%. After applying FVF, the FARs on the Linux kernel projects decreased significantly by 63.8\% to 85.0\%. For Redis, all the false alarms are identified and filtered by FVF.
}

In summary, the overall FAR of all \revise{four} baselines is \revise{76.2\%}, which is far from satisfactory and impractical.
FVF identifies \revise{233} fixed vulnerabilities in \revise{358} potential vulnerabilities, reducing the overall FAR from \revise{76.2\%} to \revise{26.6\%}. Even with the trivial hash-based approach, FVF reduces FAR from 60.3\% (32 out of 53) to 7.5\% (4 out of 53).
These results demonstrate the effectiveness of FVF in reducing false alarms in clone-based vulnerability detection.

\find{\textbf{RQ1 Result:} {The false alarm rate of the \revise{four} existing clone-based vulnerability detection approaches are high and far from satisfactory. FVF is proven to be effective in reducing false alarms.}}

\subsection{RQ2: False Predictions of FVF in Identifying False Alarms of Clone-based Vulnerability Detection Approaches}
\label{sec:rq2}

FVF aims to reduce the false alarms of clone-based vulnerability detection approaches by identifying SBP code snippets. In this RQ, we further look into the details of when FVF fails.

\subsubsection{\textbf{False Positives of FVF}}
\label{sec:false_positive}
False positives of FVF refer to real recurring vulnerabilities that are incorrectly filtered as SBP code by FVF.
This may hinder the actual vulnerabilities and is often unacceptable.
We manually verify all SBP code snippets identified in RQ1 and find no false positive case.

FVF employs an evidence-based process and adopts a conservative strategy to identify SBP code snippets. Specifically, FVF considers a vulnerable function detected by a clone-based vulnerability detection approach as SBP if and only if all code changes in the vulnerability patch log are rigorously contained in the function change log.
This ensures that the SBP identified by FVF must have been historically patched.

\subsubsection{\textbf{False Negatives of FVF}}
\label{sec:false_negative}

\begin{figure}
    \centering
    \centerline{\includegraphics[width=\linewidth]{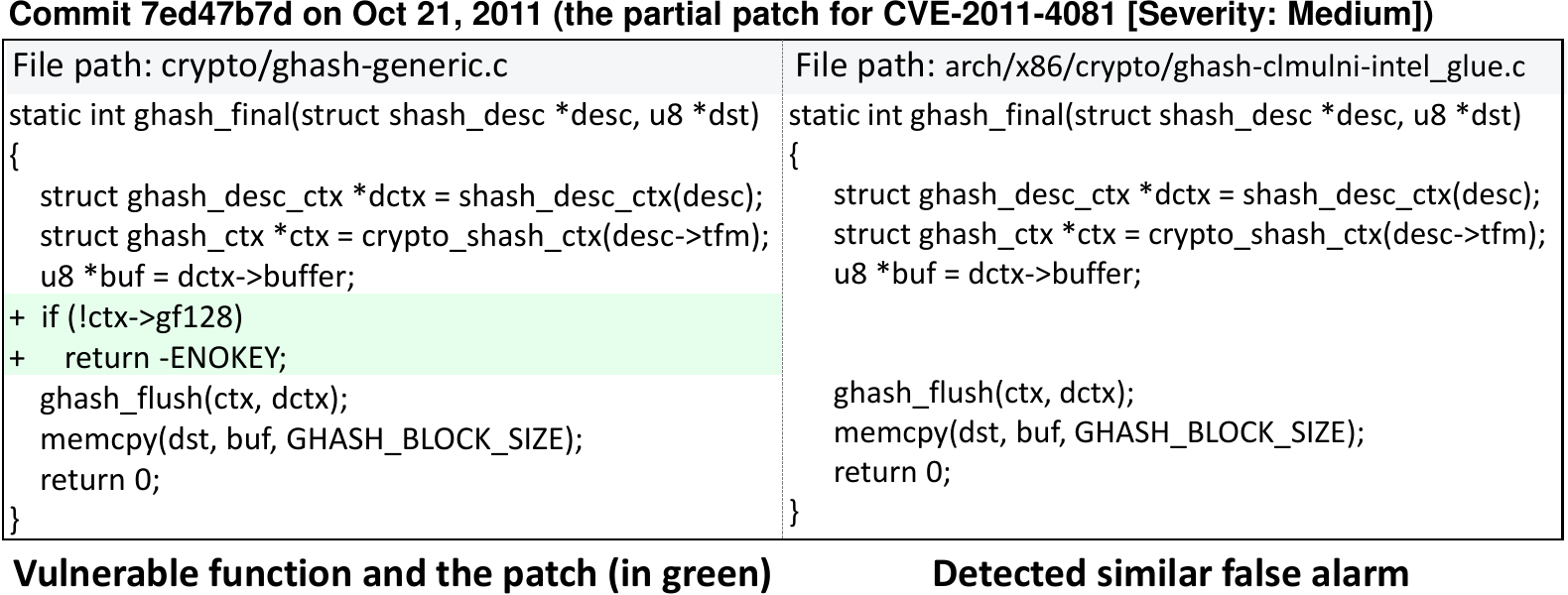}}
    \vspace{-0.2cm}
    \caption{A false alarm falls outside of SBP. The complete patch for CVE-2011-4081~\cite{CVE-2011-4081_patch} includes two identical code changes. Here we show the first.}
    \vspace{-0.5cm}
    \label{fig:discussion_fail1}
\end{figure}

\noindent False negatives are false alarms produced by the underlying clone-based vulnerability detectors but not filtered by FVF.
While FVF can significantly reduce false alarms~(specifically, the SBP code) as shown in RQ1, there are cases where FVF misses certain false alarms. In RQ1, there are 125 false alarms that FVF does not identify.
We manually conduct a qualitative study to investigate the reason why FVF fails and summarize two situations:

\noindent\textbf{\ding{182} Lack of vulnerability trigger point.}
We observe 90 cases that do not have fix behaviors in the past, however, are not vulnerable.
What makes the difference is the calling context, i.e., the contextual conditions. The contextual conditions required for the vulnerabilities are not satisfied for the detected functions, therefore, they are false alarms.

Figure~\ref{fig:discussion_fail1} shows a failed example for CVE-2011-4081~\cite{CVE-2011-4081} and the detected false alarm. The root cause of the vulnerability is a null pointer dereference due to the pointer (\texttt{ctx->gf128}) could be null. The vulnerability was fixed on Oct 21, 2011, by adding a null pointer check and returning an error code in the null case. The detected false alarm, with the same name {\tt ghash\_final}, is also in the Linux project but exists in a different file. Since the pointer is not used within that file, there is no vulnerability or need for a null check.

In such cases, contextual information, such as function calls and value flow, becomes critical in determining if a similar function is truly a vulnerability. To identify this type of false alarm, {\textit{inter-procedural analysis}} could be introduced.
However, filtering such cases is out of the scope of FVF, while we focus on false alarms caused by SBP code.

\noindent\textbf{\ding{183} Trivial clone false alarms.} There are 35 \revise{false alarms failed to be identified}, which code is neither related to the vulnerabilities nor the patches. These cases are mostly short auxiliary functions or even non-function code and are hard to associate with any vulnerabilities. All these cases are produced by ReDeBug, which constructs vulnerability signatures using partial context lines in patch diffs. Although this improves the scalability, it also introduces more false alarms.
Filtering out such false alarms falls is also out of the scope of FVF.

\find{\textbf{RQ2 Result:} {In our evaluation, FVF has no false positives and keeps a low false negative rate in reducing false alarms. {It is important to note that the few missed false alarms are mainly out of the scope of FVF.}}}

\subsection{RQ3: The Characteristics of Identified SBPs}
\label{sec:rq3}

\begin{figure}[t!]
    \centering
    \includegraphics[width=\linewidth]{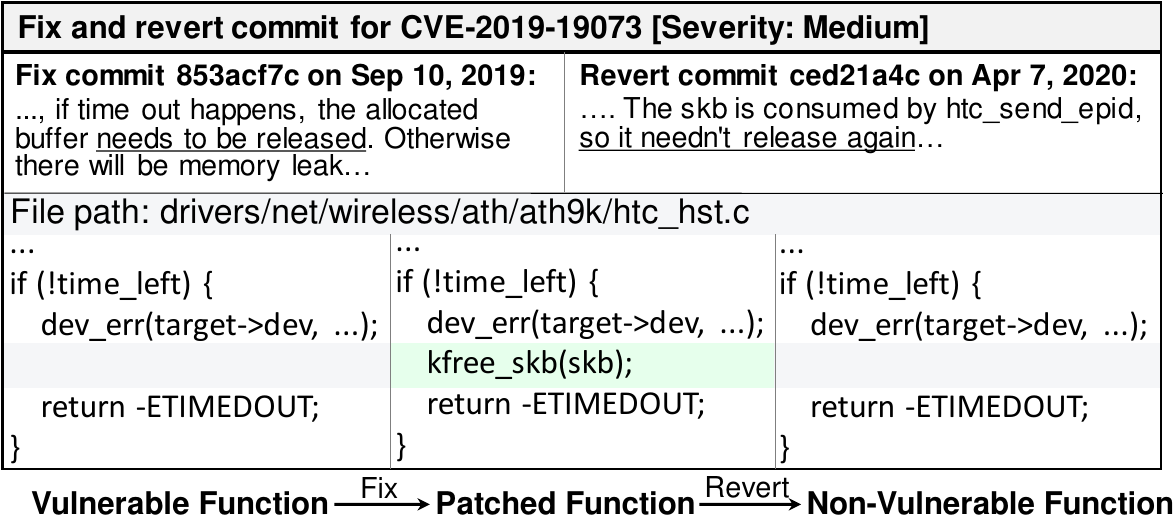}
    \caption{An example of patch reversion. Commit ced21a4c~\cite{CVE-2019-19073_revert} reverts the change made by commit 853acf7c~\cite{CVE-2019-19073_patch} -- the fix for CVE-2019-19073~\cite{CVE-2019-19073}.}
    \label{fig:rollback_type}
    \vspace{-8mm}
\end{figure}

\noindent In total, FVF identified and filtered out 238 false alarms in RQ1.
After manual verification, we confirm that all of them are already patched and should be filtered out.
In this RQ, we further analyze the characteristics of these filtered false alarms to gain empirical insights.
We categorize them into three categories: 1) Patch Reversion (126 cases), 2) Minor Difference (86 cases), and 3) Customized Patch (26 cases).

\noindent\textbf{\ding{182} Patch Reversion} refers to the cases where a vulnerable function is first fixed, but then the modification is rolled back. There are 126 patch reversion cases in total.
Reverting a commit involves undoing the change made before.
\revise{It is found to be a common operation during software development due to reasons such as unexpected software regression~\cite{tan2015relifix} and the introduction of new bugs~\cite{shimagaki2016why}.
We observe that patch reversion also occurs in the context of vulnerability fixing.}

Figure~\ref{fig:rollback_type} shows an example of a patch of CVE-2019-19073~\cite{CVE-2019-19073_patch} and the corresponding reversion.
The root cause is that, when timed out, the allocated socket buffer {\tt skb} is forgotten to be released, resulting in a memory leak. However, the patch (made on Sep 10, 2019) was reverted on Apr 7, 2020, since {\tt skb} is consumed by another function ({\tt htc\_send\_epid}). Therefore, it is incorrect to release it again.
After reverting the patch, the function is the same as the vulnerable one first reported in CVE-2019-19073~\cite{CVE-2019-19073} but is no longer vulnerable.
Most existing vulnerability detection approaches (e.g., the clone-based vulnerability detection~\cite{jang2012redebug, kim2017vuddy, xiao2020mvp}, DL-based vulnerability detection~\cite{zhou2019devign, fu2022linevul, hanif2022vulberta} fail to distinguish reverted functions from vulnerable ones, since they only rely on the information of the function itself.
This also verifies the unique advantages of our proposed FVF by considering the function change logs.

We further analyze the messages of the revert commits and identify two primary reasons for patch reversions:
1) \textit{Contextual Change}. In 28 cases, the commit messages suggest that the vulnerable version is no longer vulnerable due to a change in context.
2) \textit{Inadequate Patch}. In 18 cases, the original patch is found to be incorrect or inadequate to fix the vulnerability, and it is easier to develop a new patch from scratch rather than revise the original patch.

This phenomenon indicates the secrecy of the reverted patches. It is noteworthy that the cases of patch reversions are not rare, and the corresponding research is insufficient. So far, no baseline can handle the reverted patches.

\noindent\textbf{\ding{183} Minor Difference} refers to false alarms where a patched function is wrongly detected as vulnerable\revise{, but not due to patch reversion or customized patch}.
There are 86 such cases in total.
We analyze the difference between vulnerable and patched functions based on three metrics: lines of code~(LOC), added lines of code~(ALOC), and removed lines of code~(RLOC).
\revise{The median values are 4, 0, and 3, respectively.}
This confirms the challenges in distinguishing patched functions from vulnerable ones due to the subtle differences.

\noindent\textbf{\ding{184} Customized Patch} refers to false alarms in which developers apply the original patches from upstream but then customize the patched code to fit the downstream. After customization, the signatures of the customized functions become similar to those of the vulnerable functions, resulting in false alarms of clone-based vulnerability detection approaches.
We observe eight false alarms of the customized patch.
Due to the customization requirements, upstream patches may require additional development or refactoring, as discussed in previous studies~\cite{wang2021patchdb, tan2022understanding}.
Some existing approaches, such as ReDeBug, also consider patch information. However, these approaches may still fail to distinguish customized patches from vulnerabilities.
The reason is that most fix information is typically extracted from the mainline version, and customized patches are often not taken into consideration~\cite{wang2021patchdb}.
FVF can identify these false alarms by analyzing the code change history. By detecting the core fix behaviors, FVF can detect the function as a potential false alarm.

In conclusion, among the 238 filtered SBP-related false alarms, 52.9\% (126 cases) are patch reversion false alarms,
36.1\% (86 cases) are minor difference false alarms
 and 10.9\% (26 cases) are customized patch false alarms.
All three cloned vulnerability detection approaches fail to correctly distinguish these fixed vulnerabilities, generating a large number of false alarms.
In contrast, our proposed approach, FVF, can effectively reduce these false alarms caused by SBP.

\find{\textbf{RQ3 Result:} We categorize the 238 filtered false alarms into three categories {and provide insights of each category}: Patch Reversion, Minor Differences, and Customized Patch. {Our analysis demonstrates the unique advantages of FVF in reducing the false alarms caused by SBP code by utilizing the function change logs.}}

\section{Discussion}
\label{sec:discussion}
\noindent In this section, we further generalize FVF to large-scale real-world open-source software~(OSS) projects and construct an SBP dataset.
We aim to answer:
1) Can FVF be generalized to large-scale real-world projects?
2) What is the impact of the SBP code on deep learning-based vulnerability detection approaches?
We also discuss the prevalence of SBP code in {real-world} projects, as well as the efficiency and the programming language agnostic nature of FVF.

\subsection{Prevalence of SBP Code in Open-Source Projects}
\label{sec:dataset_construction}
\label{sec:dataset_prevalence}

\begin{figure}[t]
    \centering
    \includegraphics[width=.9\linewidth]{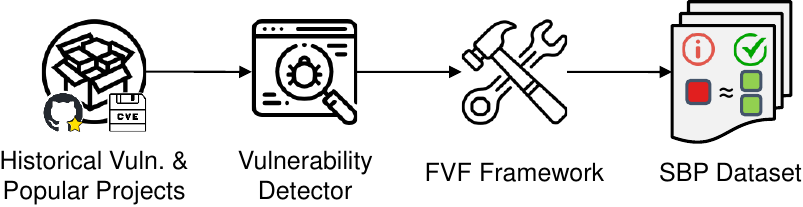}
    \vspace{-0.2cm}
    \caption{SBP Dataset Construction.}
    \vspace{-0.5cm}
    \label{fig:dataset}
\end{figure}

We leverage the same framework as in Section \ref{sec:overall_framework} to construct the SBP dataset.
Figure \ref{fig:dataset} presents an overview of the construction of the dataset.
We employ VUDDY-J, an extended version of VUDDY~\cite{kim2017vuddy} that supports Java programming language, as the vulnerability detector,
{with vulnerability source collected from MITRE~\cite{footnote_mitre},}
{to scan for vulnerable code clones on a large number of OSS projects.}
The detected potentially vulnerable code clones are further verified by FVF to identify the instances of SBP, {which forms our dataset.}

The scanned projects are selected from two sources: historically vulnerable projects from CVE and popular projects in the wild.
Historically vulnerable projects provide cases where the fixed function (e.g., backported patches) is similar to the vulnerable function.
Popular projects allow us to investigate the presence of the SBP phenomenon in the wild.
The top 100 most starred projects on GitHub for the C, C++, and Java programming languages are selected, respectively.
In total, we collected 1,081 projects for detection.
Since a project may contain different software versions in different branches, the detection is only performed on active branches with code commits within the last five years.

After deduplication, we collect
6,827 SBP functions, which are similar to 3,945 vulnerable functions associated with 2,834 vulnerabilities.
To ensure the functions included in our dataset are truly SBPs, we manually check 364 randomly sampled SBP functions (with a confidence level of 95\% and a margin of error of 5\%) and verify that all of them are patched.
This is not ideal, but manually examining all samples requires enormous human labor and is not feasible. Moreover, we have shown in RQ2 that FVF, as an evidence-based approach, achieves a precision of 100\% (i.e., no falsely identified SBPs).
It is noteworthy that all the sampled functions are confirmed as real SBP code, demonstrating the accuracy of FVF and its low false positive rate.

We assess the prevalence of SBP code in three perspectives: project, vulnerability, and function.
Out of the 1,081 OSS projects studied, 40\%~(430 projects) contain at least one instance of SBP.
Among the 8,570 vulnerabilities, 31\%~(2,647 vulnerabilities) have at least one vulnerable function with SBP variants.
Additionally, among the 22,471 vulnerable functions collected, 18\%~(3,945 vulnerable functions) have corresponding SBP variants.
These results demonstrate the significant presence of SBP, emphasizing the importance of addressing the challenge of SBP.

\subsection{Programming Language Agnostic}

The SBP dataset is only collected from C, C++, and Java projects, however,
the core design of FVF is programming language agnostic, and FVF can easily extend to other programming languages.
In FVF, generation of the \textit{function change log} and \textit{patch log} only relies on the history tracking feature of version control systems (e.g., Git and SVN), which is a general feature regardless of programming languages. The matcher module applies log comparison, which is also not limited to specific languages.
Hence, FVF is compatible with any programming language as long as the source code is managed with the standard version control system.

\subsection{Impact of SBP code on Deep Learning-based Vulnerability Detection Approaches}
\label{sec:dl_based}

DL-based vulnerability detection approaches have shown promising performance in controlled lab environments.
However, the SBP cases are not well considered in the previous work. In the SBP dataset, the vulnerable code and the SBP code share subtle differences, but the vulnerability just manifests itself in these differences.

In the experiment, we evaluate two token-based and two graph-based state-of-the-art DL-based vulnerability detection approaches with the SBP dataset.
Similar to RQ1~(Section \ref{sec:evaluation_metrics}), we employ FA and FAR as metrics.
We select the following state-of-the-art DL-based approaches:

    \noindent \textbf{LineVul}~\cite{fu2022linevul} utilizes a transformer-based language model to generate vector representations, enabling both the prediction of vulnerable functions and the localization of vulnerable lines.

    \noindent \textbf{VulBERTa}~\cite{hanif2022vulberta} utilizes RoBERTa, a transformer-based deep learning model with a custom tokenization pipeline to detect vulnerabilities. It has two variants: a convolution neural network~(V-CNN) and a multilayer perceptron~(V-MLP).

    \noindent \textbf{Devign}~\cite{zhou2019devign} utilizes a gated graph neural network~(GGNN) to learn program dependencies features on code property graphs, enabling effectively detection of function vulnerabilities.

    \noindent \textbf{IVDetect}~\cite{li2021vulnerability} utilized a sliding window technique combined with an interpretable graph neural network to detect vulnerabilities and provide fine-grained explanations. It extracts five different scale features from the code and archives state-of-the-art performance on multiple datasets.

For all techniques, we use the pre-trained model weights provided by the authors and the default parameters mentioned in the original papers.
As these deep learning models are designed mainly for C/C++, we only use the C and C++ data~(about 97\% of all).

\begin{table}[t]
    \caption{Performance of state-of-the-art DL-based vulnerability detection techniques on the SBP dataset.}
    \vspace{-0.2cm}

    \resizebox{\linewidth}{!}{
    \begin{tabular}{@{}lrrrrr@{}}
    \toprule
    \multirow{2}{*}{
        \textbf{Approach}
    }
    & \multicolumn{3}{c}{\textbf{Token-Based}}
    & \multicolumn{2}{c}{\textbf{Graph-Based}} \\
    \cmidrule(l){2-4} \cmidrule(l){5-6}
    & \textbf{LineVul} & \textbf{V-CNN} & \textbf{V-MLP} & \textbf{Devign}    & \textbf{IVDetect}   \\
    \cmidrule{1-6}
    \textbf{False Alarms}                        & 6,822            & 6,827          & 6,644          & 2,743              & 2,388               \\
    \textbf{FA Rate(\%)}                & \textbf{63.4}   & \textbf{64.4} & \textbf{63.5} & \textbf{64.9} & \textbf{62.9}   \\

    \bottomrule
    \end{tabular}
    }
    \vspace{-5mm}
    \label{tab:sota_performance}
\end{table}

The evaluation results are presented in Table \ref{tab:sota_performance}.
The average false alarm rate~(FA Rate) of all models in the SBP dataset is as high as 63.8\%.
Specifically, token-based models predict almost all SBP cases as vulnerable. For example, the VulBERTa-CNN model predicts that all SBP samples are vulnerable.
For the graph-based models, 64.9\%~(2,743) of the Devign predicted positives are SBP code, and 62.9\%~(2,388) of IVDetect predicted positives are SBP code.

The results demonstrate that the DL-based approaches also fail to distinguish the SBP code accurately and {the SBP code poses large challenges to these approaches.}
Therefore, the SBP dataset can serve as a benchmark for evaluating the sensitivity of existing vulnerability detection approaches to accurately identify real vulnerabilities among the SBP data.

\subsection{Time and Efficiency}
\label{sec:discussion_efficiency}
In this section, we discuss the efficiency and overhead of FVF.
FVF leverages the standard software version control system to retrieve \textit{function change log}s, queries \textit{patch logs} from the vulnerability feature database, and then uses a highly efficient code change log matching to match the two log sequences.
The key overhead of FVF is the process of generation of function change logs. In a large repository with numerous commits, retrieving histories for specific functions or files is usually a heavy operation, which slows down the overall process. However, without sufficient change logs, FVF may fail to filter false alarms. Therefore, the size of the retrieval window is critical for the accuracy and efficiency of the algorithm.
For this reason, we conduct an empirical analysis to determine the optimal retrieval window size.

Figure \ref{fig:backtrack_times} presents the empirical cumulative distribution function~(ECDF) of required retrieval times of filtered false alarms in RQ1, which demonstrates that all filtered false alarms can be filtered within a maximum retrieval window size of 55. Additionally, the majority~(93\%) of successful cases are filtered by retrieving only a few ($\leq$ 20) change histories. The ``hard'' cases, which require more retrievals and time, only constitute a minority~($\leq$ 7\%) of the overall cases.

Then we experiment to determine the best retrieval window size for optimal performance and efficiency.
We use the false alarm data in RQ1~(Section \ref{sec:results})
and then reapply FVF under different maximum retrieval window size settings to test its performance. We use the same metrics FA, FAR, and IR. Additionally, we also measure the elapsed time and compute the average processing time for each case.

The results, as shown in Table \ref{table:backtrack_times}, indicate that with a larger retrieval window, the improvement rate increases, while the time and the average time are longer. Considering this trade-off between efficiency and performance, we empirically recommend a maximum retrieval window size of 50.

\begin{table}[!t]
  \centering
  \caption{The time efficiency and performance under different thresholds of retrieval window size. (FA: False alarm, FAR: False alarm rate, IR: Improvement rate, Avg.: Average time.)}
  \vspace{-0.1cm}
  \resizebox{\linewidth}{!}{
    \begin{tabular}{@{}p{1.3cm}p{.9cm}p{.9cm}p{.9cm}p{.9cm}p{.9cm}@{}}
    \toprule
    \textbf{\makecell[vl]{Threshold}}
    & \textbf{\mbox{FA}}  & \textbf{\mbox{FAR(\%)}}  & \textbf{IR(\%)}  & \textbf{Time(s)} & \textbf{\makecell[vl]{Avg.(s)}}
    \\ \midrule
    \textbf{3}  &   197  &  31.9   &  68.1  &  168 & 0.27 \\
    \textbf{10} &  149   &  24.2    &  75.9   &  182 & 0.30 \\
    \textbf{20} &  129   &  20.9    &  79.1   &  192 & 0.31 \\
    \textbf{50} &  117   &  19.0    &  81.0   &  245 & 0.40 \\
    \textbf{55} &  109   &  17.7    &  82.3   &  257 & 0.42 \\
    \bottomrule
    \end{tabular}
  }
  \label{table:backtrack_times}
  \vspace{-5mm}
\end{table}

\subsection{Extensibility of FVF}
\revise{
Similar to prior works, FVF is primarily designed for single-function similar vulnerability detection.
However, FVF can also be extended to vulnerabilities across multiple functions (i.e., the inter-procedural vulnerabilities).
Since the core design of FVF is to utilize the fix behavior, to detect similar inter-procedural vulnerabilities, FVF can further utilize all fix behaviors of relevant vulnerable functions to generate the multi-function patch log for the inter-procedural vulnerabilities, which we plan as future work.
}

\section{Threats To Validity}
\label{sec:threats}

\noindent \textbf{Internal validity}.
Threats to internal validity are associated with bias and errors in the experiment.
One potential threat is the absence of a ground truth in RQ1, where the predicted results are manually analyzed, potentially introducing bias.
To mitigate the bias, we engage security professionals with at least three years of experience in software security to conduct the manual analysis.
{Another potential threat is the accuracy of the SBP dataset.
To mitigate the threat, we conduct a manual verification process on a random sample of 364 SBP functions, with a confidence level of 95\%.}
\revise{Another threat is the comprehensiveness of our vulnerability feature database.
Vulnerabilities can be fixed through custom fixes that are different from the records in the CVE database. To mitigate the threat, we gather vulnerability fix commits from various public vulnerability databases and datasets to enhance our database.
}

\noindent \textbf{External validity}.
Threats to external validity are related to the quality of vulnerability data.
Previous research reveals that developers often group multiple changes into a single commit, resulting in a tangled code change~\cite{herzig13tangleimpact}, which produces large noise and bias.
\revise{In FVF, we remove VFCs that modify more than 10 files or only modify comments, white spaces, or test/log files to mitigate the threat. This could still be improved using commit untangling techniques such as~\cite{wang2019untangling}.}
\revise{Another threat is associated with the selection of tested projects. In this study, we select projects with extensive branches and forks as our targets, but there may be better factors, we encourage future studies to investigate more factors in selection.}

\begin{figure}[!t]
\begin{center}
\includegraphics[width=.95\linewidth]{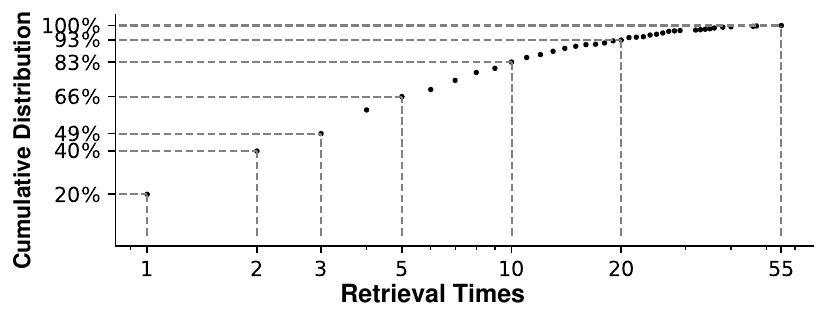}
\vspace{-0.4cm}
\caption{ECDF of required retrieval times of filtered FAs.}
\vspace{-0.7cm}
\label{fig:backtrack_times}
\end{center}
\end{figure}

\section{Related Work}
\label{sec:related_work}

\noindent\textbf{Clone-Based Vulnerability Detection.}
Various approaches have been proposed for detecting recurring or similar vulnerabilities~\cite{li2006cp, pham2010detection, jang2012redebug, li2016vulpecker, kim2017vuddy, xiao2020mvp, cui2020vuldetector, salimi2022vulslicer, woo2022movery, bowman2020vgraph}.
For example, Jiang et al.~\cite{jiang2007deckard} and Kim et al.~\cite{kim2017vuddy} consider code as token sequences to detect vulnerability clones.
\revise{Bowman et al.~\cite{bowman2020vgraph} leverage code property graphs to enhance the robustness of the matching algorithm against code modifications.}
\revise{Xiao et al.~\cite{xiao2020mvp} consider code snippets that match the vulnerable version and not match the fixed version as recurring vulnerabilities.}
\revise{Kang et al.~\cite{kang2022tracer} use taint analysis traces to detect recurring vulnerabilities but are limited to taint-style C/C++ vulnerabilities and do not incorporate code change history.}
Woo et al.~\cite{woo2022movery} consider the oldest, disclosed, and patched versions of the vulnerable function to generate more robust signatures to find modified vulnerable code clones.
\revise{However, these works focus only on the code snippets and overlook the extensive information contained in the change histories, resulting in many false alarms in clone-based vulnerability detection. In contrast, our work pays attention to the fix reversions of vulnerabilities in real-world scenarios, utilizes a broader history to make more comprehensive decisions, and significantly reduces the number of false alarms.}

\noindent\textbf{Deep Learning-Based Vulnerability Detection.}
Deep learning-based approaches have made periodic achievements in vulnerability detection~\cite{li2018vuldeepecker, zhou2019devign, li2021vulnerability, zou2021interpreting, chakraborty2021reveal, li2021sysevr, li2021vuldeelocator, hanif2022vulberta, fu2022linevul, shi2019vulnerable, wang2020combining, wang2024combining}. These works typically utilize deep neural networks to learn vulnerable patterns from various forms of code representation, such as lexical tokens~\cite{fu2022linevul, hanif2022vulberta}, program dependence graphs~\cite{li2018vuldeepecker, li2021sysevr, li2021vuldeelocator, zou2021interpreting, wang2020combining}, and a mixture of multiple representations~\cite{zhou2019devign, chakraborty2021reveal, li2021vulnerability, wang2024combining}.
LineVul~\cite{fu2022linevul} employs a transformer-based architecture to generate vulnerability representations for line-level vulnerability detection.
Devign~\cite{zhou2019devign} introduces GGNN to learn data and control dependencies features of vulnerable code.
\revise{Shi et at.~\cite{shi2019vulnerable} train a graph convolutional network~(GCN) on historically vulnerable functions and correlations to detect cloned vulnerabilities in downstream operating system distributions.}
\revise{Concoction~\cite{wang2024combining} extracts both static and dynamic features of the code and uses a bidirectional Transformer network for vulnerability detection.}
However, these works often struggle to distinguish subtle differences between vulnerabilities and their corresponding fixed versions and therefore showed poor performance on SBP cases~(see experimental results in Section \ref{sec:dl_based}).

\noindent\textbf{Other Kinds of Vulnerability Detection Approaches.}
Various works and techniques are also available in the field of vulnerability detection.
Static analysis-based approaches~\cite{clang_static_analyzer, cppcheck, flawfinder, lu2019detecting} detect vulnerabilities through induction of possible variable values.
Fuzzing-based approaches~\cite{cha2015program, aflplusplus, gross2023fuzzilli} aim to crash the program using random input to uncover vulnerabilities.
Symbolic execution-based approaches~\cite{anand2007jpf, cadar2008klee, ramos2015under} explore feasible execution paths by symbolizing variables to assist testing.
Compared to these techniques, FVF is lightweight and does not require compiling or executing the code, which is a costly operation.
Furthermore, FVF can also be complemented with these techniques. After eliminating SBP-style false alarms, other techniques can be applied for further validation.

\section{Conclusion}
\label{sec:conclusion}

In this paper, we focus on the SBP phenomenon in vulnerability detection.
We propose a new {programming language agnostic} framework, FVF, to identify SBP cases and reduce false alarms in vulnerability detection.
Our evaluation conducted with four cloned-based vulnerability detection tools and across nine versions of the Linux and the Redis project demonstrates that FVF can significantly reduce false alarm rates.

We further apply FVF to 1,081 real-world projects and construct a real-world SBP dataset containing 6,827 SBP functions.
Using the dataset, we demonstrate the ineffectiveness of state-of-the-art DL-based vulnerability detection approaches in distinguishing SBP data.
The dataset can help developers make a more realistic evaluation of existing vulnerability detection approaches and also paves the way for further exploration of real-world SBP scenarios.

\section*{Acknowledgments}

This research is supported by the Ningbo Natural Science Foundation (No. 2023J292).

\balance

\bibliographystyle{IEEEtran}
\bibliography{IEEEabrv,sample-base}

\end{document}